\documentclass[%
prb,twocolumn,
superscriptaddress,
amsmath,amssymb,
 aps,
]{revtex4-2}

\usepackage{graphicx}
\usepackage{dcolumn}
\usepackage{bm}
\usepackage[mathlines]{lineno}

\usepackage{physics}
\usepackage{amsmath}
\usepackage{blkarray}
\usepackage{xcolor}
\usepackage{natbib}
\usepackage[pagebackref=false,colorlinks,linkcolor=blue,citecolor=blue,urlcolor=blue]{hyperref}
\usepackage{soul}

\begin{document}

\preprint{APS/123-QED}

\title{Does Topology Enhance Thermoelectric Efficiency?\\A Case Study in  Bismuthene}

\author{Muhammad Gaffar}
\affiliation{%
School of Physics and Astronomy, Shanghai Jiao Tong University, Shanghai 200240, China.
}
\affiliation{%
Tsung-Dao  Lee  Institute,  Shanghai  200240,  China.
}

\author{Sasfan Arman Wella}
\affiliation{
Research Center for Physics, Indonesian Institute of Sciences (LIPI), Tangerang Selatan 15314, Indonesia.
}%

\author{Eddwi Hesky Hasdeo}
\email{eddw001@lipi.go.id}
\affiliation{
Research Center for Physics, Indonesian Institute of Sciences (LIPI), Tangerang Selatan 15314, Indonesia.
}%
\affiliation{
Department of Physics and Material Science, University of Luxembourg, L-1511 Luxembourg.}

\date{\today}

\begin{abstract}

Two-dimensional (2D) bismuth (Bi) layer, known as bismuthene, exhibits $Z2$ topological bulk states due to large spin-orbit coupling that inverts the bands.
Using the tight-binding method, we calculate the band structure of buckled bismuthene to understand its topological and trivial phases. 
We determine the thermoelectric properties for some considered phases, incorporating the edge states contribution, by using the linearized Boltzmann transport equation (BTE) with a constant relaxation time approximation. 
It is shown that the thermoelectric figure of merit, $ZT$, actually drops in undoped topological bismuthene due to the edge effects.
Surprisingly, the topological edge states enhance $ZT$ at large doping with the Fermi energy near the bottom of bulk bands when bismuthene is nearly metallic. 
\end{abstract}

\maketitle


\section{\label{sec:intro} Introduction}
%

Thermoelectric (TE) effect is useful to convert waste heat into electricity. 
Even though the effect has been known for a century, the high-performance TE materials are still highly desired. 
TE performance is quantified by the dimensionless figure of merit ($ZT = \sigma S^2 T/\kappa$)~\cite{goldsmid2010introduction}, where, $\sigma$ is electrical conductivity, $S$ is the Seebeck coefficient that gives the amount ratio of generated gradient of potential to a given gradient of temperature, and $\kappa=\kappa_e+\kappa_{\rm ph}$ is thermal conductivity consisting of electronic and phonon contributions. 
The development in high-performance TE materials, unfortunately, are limited by the interdependence of  the TE quantities that constitute the \textit{ZT}.

Since the beginning of TE research, Bi-based materials, such as $\rm Bi_2Se_3$ and $\rm Bi_2Te_3$, have been of prime importance~\cite{zhang2009topological, bi2te3, zhang2011band, xie2010identifying, muchler2013topological, Chen178, hermannbi2se3, hongliangbi2te3}.
Having small energy (band) gaps $\sim0.1-0.2$ eV~\cite{ryu2016computational}, these Bi-based crystals satisfy the optimal band gap for TEs set by Sofo and Mahan~\cite{mahan1996best}.  
It turns out that large spin-orbit coupling (SOC) in Bi inverts the valence and conduction bands and consequently reduces the band gap~\cite{Bernevig1757}. The SOC gives rise to the spin-Seebeck effect~\cite{bi2te3} and, importantly, renders the Bi-based crystals topological insulators (TIs) through the band inversion~\cite{muchler2013topological, TIcolloquium, Bernevig1757, kaneinversion, hsieh2008topological}. 
It naturally raises an interesting question of whether the electronic properties of TIs can affect TE performance.
A substantial number of theoretical and experimental works thus far attempt to explore topological insulators for TE applications~\cite{xu2017topological,fu2020topological,baldomir2019behind,takahashi2012thermoelectric,xuTITE,ghaemiTITE,liang2016maximizing,mavrokefalos2009thermoelectric,guo2016tuning,zhangthinfilms,tang2015thermoelectric,teweldebrhan2010exfoliation}.

Basically, TIs are similar to ordinary insulators with bulk band gaps but possess gapless edge states. 
From TE transport point of view, these topological edge states serve as metallic channels.
As a result, they will increase $\sigma$ and $\kappa_e$ in the gap, while $S$ is decreased. 
Accordingly, it is still unclear TIs will always improve the TE properties. 

\begin{figure}[b]
\includegraphics[width=.95\columnwidth]{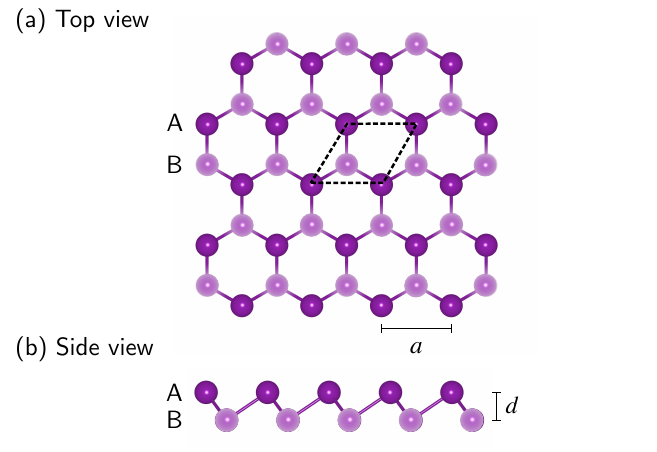}
\caption{Top (a) and side (b) views of monolayer buckled bismuthene. The color tone symbolizes different sublattices, A and B, with $d$ as the buckling height, i.e. the difference in height between sublattice A and B.}
\label{fig:structure}
\end{figure}

Thin films of Bi crystal have been actively investigated to explore its topological insulator properties~\cite{thinBi1,thinBi2,thinbi3,thinbi4}. We focus on hexagonal and buckled layered structure ($b$-Bi), since it is one of the most stable structure of two dimensional bismuth~\cite{BiBandgap}, and the thin film can be generated from 3D Bi(111) surface~\cite{iwasaki1997structural}. The monolayer structure of  buckled bismuthene is shown in Fig.~\ref{fig:structure}. 
The difference in height of two neighboring Bi atoms allows us to tune the band gap by varying the on-site staggered potential through perpendicular electric field. 
%
%
Being a 2D material, bismuthene also enjoys tunable Fermi energy (doping) simply by adjusting the gate voltage. 
Similar to 3D Bi-based crystals, bismuthene is also reported as a TI~\cite{onsitesoc}. 

To develop a coherent picture of the relation between topology and TE performance, we perform electronic structure calculations of bismuthene using the tight-binding (TB) method.
Using this method, we are able to easily change the relevant parameters for analysis.
By varying the staggered potential and on-site SOC, we can map the topological and trivial phases of bismuthene.
Using the linearized Boltzmann transport equation, we calculate TE quantities including the edge transport.
It turns out that the topological edge states decrease $ZT$ in the band gap but significantly increase it at high doping close to the metallic regime.
The reason for this behavior is that Seebeck coefficient of the edge is proportional to the doping level. 
We also highlight the enhancement of $ZT$ due to trivial edge states.
This work sheds light toward TE designs using large SOC and topological materials. 
The computational techniques developed in this work can also be applied to most of monolayer TIs.

\section{\label{sec:method} Methods}

All tight-binding calculations, to construct and solve electronic structures of bismuthene, are carried out by using {\sc{PythTB}} package~\cite{pythtb}.
The tight-binding parameters listed in Table~\ref{tab:table1} are extracted by fitting to band structures produced by the first-principles calculations. 
In order to calculate topological invariant, {\sc{Z2Pack}}~\cite{Soluyanov2011,Gresch2017} is utilized.
We explore the topological phase of the system along with its TE enhancement through bulk and edge contributions. 
The SOC is taken into account since it plays an important role to break the energy degeneracy and to produce TI phase~\cite{Kane2005}.

\subsection{\label{sec:models} Tight-binding Model of  Bismuthene}

As shown in Fig.~\ref{fig:structure}, monolayer bismuthene considered in this work is a 2D allotrope of Bi with a graphene-like structure.
The unit cell comprises two Bi atoms, with a lattice constant $a = $ 4.37 \AA.
Contrary to the flat structure of graphene, bismuthene has a periodic buckled structure with a buckling height of $d = $ 1.71 \AA.

The main contributions to the low energy bands of bismuthene come from $s$- and $p$-orbitals, while the minor contribution from $d$-orbital can be neglected to simplify our model~\cite{TBHamiltonian2015,onsitesoc}.
The TB Hamiltonian~\cite{TBHamiltonian2015} is constructed by $16\times 16$ matrix elements, consisting of $4$ atomic orbitals, $2$ sublattices, and $2$ spins. 
Considering up to the second-nearest-neighbor interactions, the TB Hamiltonian is expressed by:

\begin{table}[tb]
\caption{Tight-binding parameters for monolayer buckeld bismuthene. Parameter is given in eV.}
\label{tab:table1}
\begin{ruledtabular}
\begin{tabular}{c l r c l r c}
& Parameter & Value && Parameter & Value &\\ 
\hline
&$\varepsilon_{\rm{s}}$ & $-$7.8900 &&$\varepsilon_{\rm{p}}$ & $-$0.5630 &\\
&$V^1_{\rm{ss}}$          & $-$0.8515 &&$V^2_{\rm{ss}}$         & $-$0.0186 &\\
&$V^1_{\rm{sp}}$          &    1.3100 &&$V^2_{\rm{sp}}$         & $-$0.0200 &\\
&$V^1_{\rm{pp\sigma}}$    & $-$1.6400 &&$V^2_{\rm{pp\sigma}}$   &    0.5350 &\\
&$V^1_{\rm{pp\pi}}$       &    0.3940 &&$V^2_{\rm{pp\pi}}$      & $-$0.1090 &\\ 
\end{tabular}
\end{ruledtabular}
\end{table}

\begin{equation}
\mathcal{H} = \mathcal{H}_0 + \mathcal{H}_{\rm{hopping}} + \mathcal{H}_{\rm{OS}} + \mathcal{H}_{v} ,
\label{eq:hamiltonian}
\end{equation}
where $\mathcal{H}_0$, $\mathcal{H}_{\textrm{hopping}}$, $\mathcal{H}_{\textrm{OS}}$, and $\mathcal{H}_{v}$ are Hamiltonians related to the on-site energy, the hopping energy, the on-site SOC, and staggered sublattice potential, respectively.

The first term in Eq.~\ref{eq:hamiltonian}, $\mathcal{H}_0$ is the on-site energy defined by
\begin{displaymath}
\mathcal{H}_0 = \sum_{j} [\varepsilon_{\rm{s}}\: a^\dagger(\textbf{R}_{j})a(\textbf{R}_{j}) + \varepsilon_{\rm{p}}\:\mathbf{b}^\dagger(\textbf{R}_{j}) \cdot \mathbf{b}(\textbf{R}_{j})].
\end{displaymath}
Here, $a^\dagger$ ($a$) is creation (annihilation) operator for $s$-orbital in lattice site $\textbf{R}_{j}$; $\bm{b}^\dagger = (b_x^\dagger,b_y^\dagger,b_z^\dagger)$ and $\bm{b} = (b_x,b_y,b_z)$ are creation and annihilation operators for three $p$-orbitals in lattice site $\textbf{R}_{j}$, respectively; and $\varepsilon_{\rm{s}}$ ($\varepsilon_{\rm{p}}$) is the on-site energy of the $s$-($p$-)orbital.  

The second term, $\mathcal{H}_{\rm{hopping}}$ corresponds to the hopping energy, including the first- ($m$=1) and the second-nearest-neighbor ($m$=2) hoppings, where the two atomic distance $\mathbf{R}_{j,k}$ are adjusted accordingly. 
The $sp$-hybridization effect is included for both hopping terms.
This hopping energy $\mathcal{H}_{\rm{hopping}}$ is expressed by
\begin{displaymath}
\mathcal{H}_{\rm{hopping}} = \sum_{\left\langle j,k \right\rangle,m} [h^m_{\rm{s}}(j,k) + h^m_{\rm{p}}(j,k) + h^m_{\rm{sp}}(j,k) + h.c.],
\end{displaymath}
with
\begin{eqnarray}
    h^{m}_{\rm{s}}({j},{k}) &=& V^{m}_{\rm{ss}}\:a^\dagger(\textbf{R}_{j})\:a(\textbf{R}_{k}),\notag\\
    h^{m}_{\rm{p}}({j},{k}) &=& V^{m}_{\rm{pp\sigma}}\:\left[{\textbf{e}_{jk}\cdot\textbf{b}^\dagger\left(\textbf{R}_{j}\right)}\right]\left[{\textbf{e}_{jk}\cdot\textbf{b}\left(\textbf{R}_{k}\right)}\right]\notag\\
    &+& V^{m}_{\rm{pp\pi}}\:\left\lbrace{\mathbf{b}^\dagger (\textbf{R}_{j})\cdot\mathbf{b}(\textbf{R}_{k})}\right.-\notag\\
    && \left.\left[{\textbf{e}_{jk}\cdot\textbf{b}^\dagger\left(\textbf{R}_{j}\right)}\right]\left[{\textbf{e}_{jk}\cdot\textbf{b}\left(\textbf{R}_{k}\right)}\right]\right\rbrace,\notag\\
    h^{m}_{\rm{sp}}({j},{k}) &=& V^{m}_{\rm{sp}}\:a^\dagger(\textbf{R}_{j})\:\left[{\textbf{e}_{jk}\cdot\textbf{b}\left(\textbf{R}_{k}\right)}\right],\notag\\
    \textbf{e}_{jk} &=& \dfrac{\textbf{R}_{j}-\textbf{R}_{k}}{\abs{\textbf{R}_{j}-\textbf{R}_{k}}},\notag
\end{eqnarray}
where, $h_{\rm{s}}$, $h_{\rm{p}}$, $h_{\rm{sp}}$ are hopping between two $s$-orbitals, hopping between two $p$-orbitals, and hopping between $s$- and $p$-orbitals. 
The associated hopping integrals are denoted by $V_{\rm{ss}}$, $V_{\rm{pp\sigma}}$, $V_{\rm{pp\pi}}$, and $V_{\rm{sp}}$.

The $\mathcal{H}_{\textrm{OS}}$, as the third term in Eq.~\ref{eq:hamiltonian}, represents the on-site SOC, induced by the relativistic effects.
The effect of Rashba is insignificant compared to dominant on-site SOC, and further it can be neglected ~\cite{onsitesoc}.
The $\mathcal{H}_{\rm{OS}}$ matrix is given by
\begin{align}
\mathcal{H}_{\rm{OS}} = \lambda_{\rm{OS}}\: \
\begin{blockarray}{ccccccc}
p_{\rm{x}},\uparrow & 
p_{\rm{y}},\uparrow & 
p_{\rm{z}},\uparrow & 
p_{\rm{x}},\downarrow & 
p_{\rm{y}},\downarrow & 
p_{\rm{z}},\downarrow &  \\
\begin{block}{(cccccc) c}
   0 &  i & 0 & 0 &  0 & -1 & \:\:p_{\rm{x}},\uparrow \\
  -i &  0 & 0 & 0 &  0 &  i & \:\:p_{\rm{y}},\uparrow \\
   0 &  0 & 0 & 1 & -i &  0 & \:\:p_{\rm{z}},\uparrow \\
   0 &  0 & 1 & 0 & -i &  0 & \:\:p_{\rm{x}},\downarrow \\
   0 &  0 & i & i &  0 &  0 & \:\:p_{\rm{y}},\downarrow \\
  -1 & -i & 0 & 0 &  0 &  0 & \:\:p_{\rm{z}},\downarrow \notag\\
\end{block}
\end{blockarray}.
\end{align}
Here, $i$ is an imaginary number, and the on-site SOC parameter used in this work is given by $\lambda_{\rm{OS}} = -0.63$ eV.

The last term in Eq.~\eqref{eq:hamiltonian}, $ \mathcal{H}_{v}$ is the staggered sublattice potential, controlled by a given electric field perpendicular to a buckled surface. The $\mathcal{H}_{v}$ can be written as
\begin{displaymath}
\mathcal{H}_{v} = \lambda_v \sum_j \xi_l \left[ a^\dagger(\textbf{R}_{j})a(\textbf{R}_{j}) + \mathbf{b}^\dagger(\textbf{R}_{j})\cdot\mathbf{b}(\textbf{R}_{j})\right],
\end{displaymath}
where, $\lambda_v$ is the staggered potential parameter, and $\xi_l=\pm 1$ for different sublattice.

\begin{figure*}[htp]
\includegraphics[width=\textwidth]{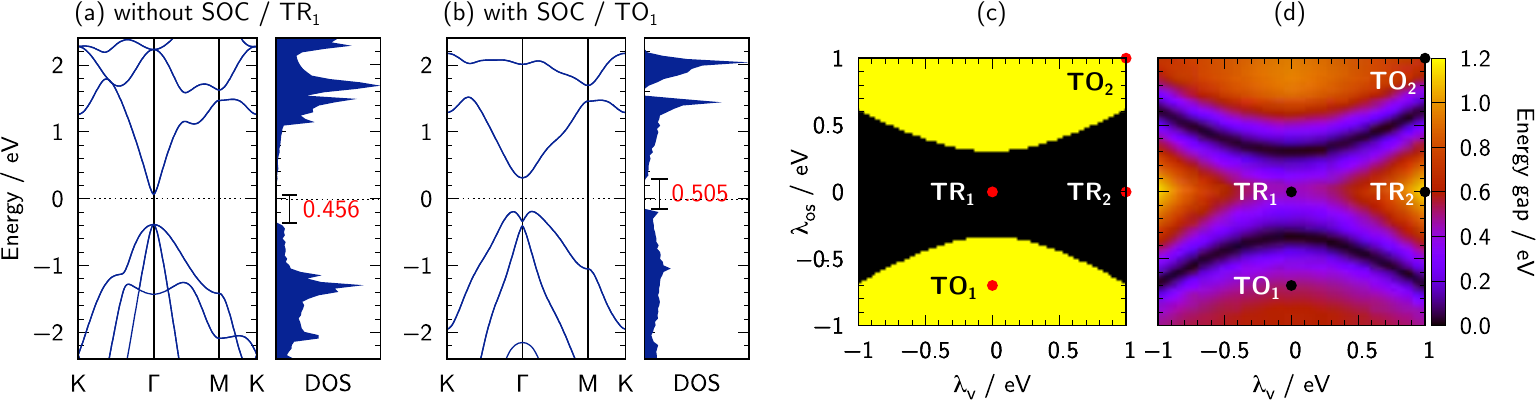}
\caption{\label{fig:bandstructure} Band structure of bismuthene (a)  without SOC, and (b) with SOC. (c) Phase diagram of bismuthene as a function of SOC coupling $\lambda_{\rm{OS}}$ and staggered potential $\lambda_{\rm{v}}$. The yellow (black) region is for topological (trivial) states. Four points (red dots) at phase diagram are sampled as representative of two topological (TO) and two trivial (TR) states for thermoelectric calculation. (d) The corresponding band gap as a function of $\lambda_{\rm os}$ and $\lambda_{\rm v}$.}
\end{figure*}

\subsection{Topological Invariant}

Topological invariant is important to predict the presence of the gapless edge state(s) in a bulk system without necessarily calculating its edge states. 
Here, topology of bismuthene is characterized by $Z2$ invariant using parity analysis, implemented in {\sc{Z2Pack}}~\cite{Soluyanov2011,Gresch2017}.
$Z2$=1 ($Z2$=0) corresponds to a topological (trivial) state.
To gain insight into the importance of topology for designing high-performance TE devices, we evaluate here the topological invariant of bismuthene by varying the SOC parameter and staggered atomic potential.

\subsection{\label{sec:citeref} Thermoelectric Transport Coefficient}

Having obtained the electronic structures from TB calculations, we then evaluate the associated TE properties by using BTE.
Deriving from BTE, we have a useful formula to calculate TE coefficient, called TE kernels, defined as
\begin{equation}
    \mathcal{L}_n^j = \int dE\; T_j(E) (E - \mu)^n \left(-\frac{\partial f(E)}{\partial E}\right), 
    \label{eq:kernel}
\end{equation}
with
\begin{displaymath}
    T_j(E) = \sum_\mathbf{k} |\mathbf{v}_{j,\mathbf{k}}|^2\tau_j(E)\delta(E - E_{j,\mathbf{k}}),
\end{displaymath}
where, ($j=$ bulk or edge) is the channel index; $f(E)$ is the Fermi-Dirac distribution; $\mu$ is the chemical potential; $T(E)$ is the transport distribution function; $\mathbf{v}_{j,\mathbf{k}} = \frac{1}{\hbar} \nabla_\mathbf{k} E_{j,\mathbf{k}}$ is a group velocity, obtained from edge or bulk band structure; and $\tau_j(E)$ is relaxation time, assumed here to be a constant ($\tau_j(E)=\tau$). 

\begin{figure}[htp]
\includegraphics[width=\columnwidth]{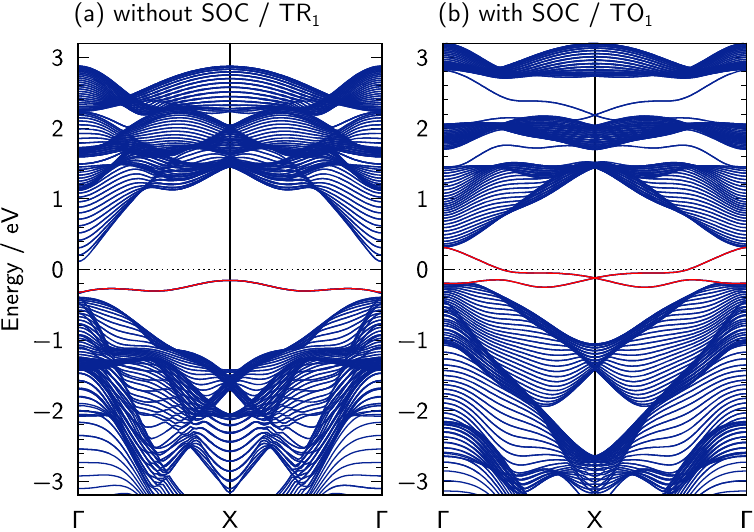}
\caption{The band structure of zigzag bismuthene nanoribbon  (a) without SOC, and (b) with SOC. Red lines indicate the edge states.}
\label{fig:edgebandstructure}
\end{figure}

In order to calculate $\mathcal{L}_n^{\rm{edge}}$, we use band structure of bismuthene nanoribbon (see Fig.~\ref{fig:edgebandstructure}) and bound the integral of Eq.~\eqref{eq:kernel} only along the bulk band gap. In present work, we use dual scattering time: (i) for topological edge states, $\tau_\mathrm{edge} \gg \tau_\mathrm{bulk}$, accounting for insensitive impurity and long-range scatterings caused by topological protection; and (ii) for trivial edge states, $\tau_\mathrm{edge}=\tau_\mathrm{bulk}$.

Combining the contribution of bulk and edge states, we have the total TE kernel, written as
\begin{align*}
    \mathcal{L}_n = \mathcal{L}_n^{\textrm{bulk}} + \mathcal{L}_n^{\textrm{edge}} . 
\end{align*}
The TE coefficient is given by
\begin{equation}
    \sigma = e^{2}\mathcal{L}_0,\:\:S = -\frac{1}{eT} \frac{\mathcal{L}_1}{\mathcal{L}_0},\:\:\kappa_e = \frac{1}{T}\frac{\mathcal{L}_0\mathcal{L}_2-\mathcal{L}^2_1}{\mathcal{L}_0},
\end{equation}
where $\sigma$ is electrical conductivity, $S$ is Seebeck coefficient, and $\kappa_e$ is electrical thermal conductivity. 
The TE performance is then characterized by calculating the power factor and figure-of-merit as follows
\begin{equation}
    \text{PF} = S^2\sigma, \quad\text{and}\quad ZT = \frac{S^2\sigma}{\kappa_e + \kappa_{ph}}T,
\end{equation}
where, $\kappa_{ph}$ is the phonon thermal conductivity. Here, we use $\kappa_{ph}$ obtained from molecular dynamics simulation~\cite{tauandkappa} and treat it as a constant.

\section{Results and Discussion}

\subsection{Band Structure}

We first consider the effect of SOC on the band structure of bismuthene.
Figures \ref{fig:bandstructure}(a) and (b) show the band structures of bismuthene without and with SOC, respectively. 
Without the SOC effect, bismuthene has a direct band gap of 0.456 eV at the $\Gamma$-point. 
This gap is quite wider than that of 3D Bi crystal.
It is confirmed that the absence of interlayer interaction will open the band gap~\cite{BiBandgap}.
With the inclusion of the SOC effect, we found that the band gap is 0.505 eV. 
Moreover, there is band inversion splitting at $\Gamma$-point, making the band gap now becomes indirect.
This band inversion gives us an indication of the existence of topological invariant states in the system.

We confirm the existence of topological states by calculating the phase diagram of the $Z2$ topological order in bismuthene as shown in Fig. \ref{fig:bandstructure}(c). At $\lambda_\mathrm{v}=0$, we see that the phase transition from trivial states (TR) with $Z_2 = 0$ to topological states (TO) with $Z_2 = 1$ is explained by the increase of absolute value of SOC. This indicates that SOC induces topological states from trivial states, while the staggered potential does the opposite, i.e. increasing the trivial band gap. At non-zero $\lambda_\mathrm{v}$ trivial area (black region) expands indicating the increase of trivial gap thus it requires larger $\lambda_\mathrm{OS}$ to make bismuthene topological.

In Fig.~\ref{fig:bandstructure}(d), we show the band gap evolution as a function of  SOC constant and staggered potential. In the absence of SOC, we see a clear band gap opening as the absolute value of staggered potential is increased (from TR$_1$ to TR$_2$). In the absence of staggered potential, the increasing absolute value of SOC will first close the band gap, and then opening the inverted band gap makes bismuthene topological. This indicates that topological transition (from TR$_1$ to TO$_1$) is simply due to band inversion caused by the SOC effect. After it experiences band inversion, the band gap will keep increasing, as the value of SOC and staggered potential increases (TO$_2$ phase).


\begin{figure}[tb]
\includegraphics[width=.95\columnwidth]{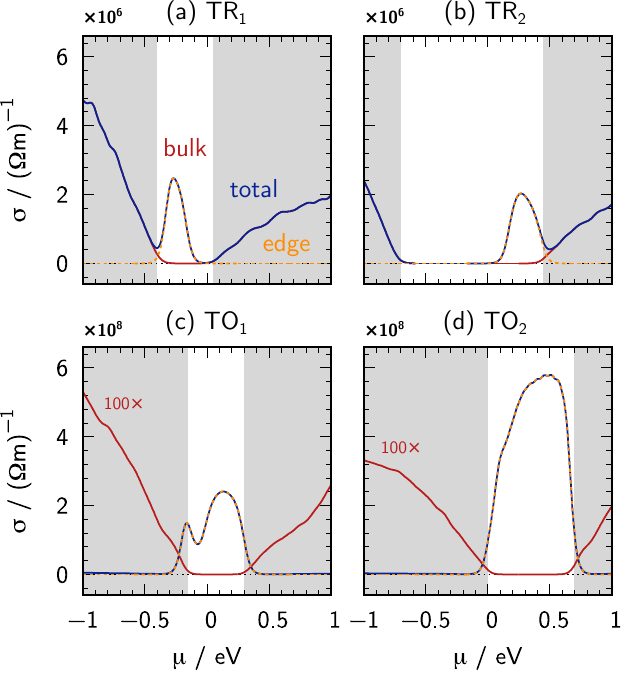}
\caption{ Electrical conductivity $\sigma$ of trivial (TR) phases (a) and (b) and topological (TO) phases (c) and (d) as a function of the Fermi energy $\mu$. The yellow, red and blue lines respectively represent edge, bulk, and total contributions. The corresponding $\lambda_\mathrm{v}$ and $\lambda_\mathrm{OS}$ parameters for all figures refer to Fig.~\ref{fig:bandstructure}(c). Note that the multiplicators in (a) and (b) are $\times 10^6$ while in (c) and (d) are $\times 10^8$. In (c) and (d), the bulk contribution is deliberately multiplied by 100 times to show the $\mu$ dependence. The gray (white) area indicates the continuum bulk (band gap) states.  }
\label{fig:sigma}
\end{figure}

\begin{figure}[tb]
\includegraphics[width=.95\columnwidth]{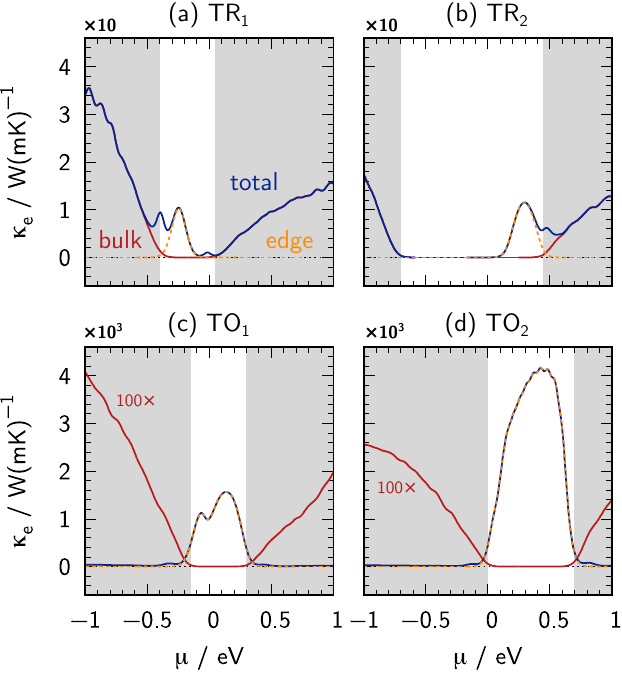}
\caption{\label{fig:kappae} Electron thermal conductivity $\kappa_e$ of trivial (TR) phases (a) and (b) and topological (TO) phases (c) and (d) as a function of the Fermi energy $\mu$. The yellow, red and blue lines respectively represent edge, bulk, and total contributions. The corresponding $\lambda_\mathrm{v}$ and $\lambda_\mathrm{OS}$ parameters for all figures refer to Fig.~\ref{fig:bandstructure}(c). Note that the multiplicators in (a) and (b) are $\times 10$ while in (c) and (d) are $\times 10^3$. In (c) and (d), the bulk contribution is deliberately multiplied by 100 times to show the $\mu$ dependence. The Gray (white) area indicates the continuum bulk (band gap) states. }
\end{figure}

Topological states also imply the existence of gapless carrier at the edge of bismuthene. 
To understand better the role of edge state in the bandstructure, we construct bismuthene as nanoribbon with zigzag boundary (30 lattice constant wide). As shown in Fig. \ref{fig:edgebandstructure}(a),  the edge dispersion of TR$_1$ state remains gapped while the one of TO$_2$ state in Fig.~\ref{fig:edgebandstructure}(b) becomes gapless. 
We need to emphasize that the cause of the transition is different from the Kane-Mele quantum spin Hall mechanism of graphene, whose transition originates from the SOC effect between the next-nearest neighbour, and the existence of non-zero Rashba SOC of $p_z$ orbital. By contrast, in bismuthene, the low-energy physics is driven by huge on-site SOC of $p_x$ and $p_y$ orbital.

\subsection{Thermoelectric Properties}

\begin{figure}[tb]
\includegraphics[width=.95\columnwidth]{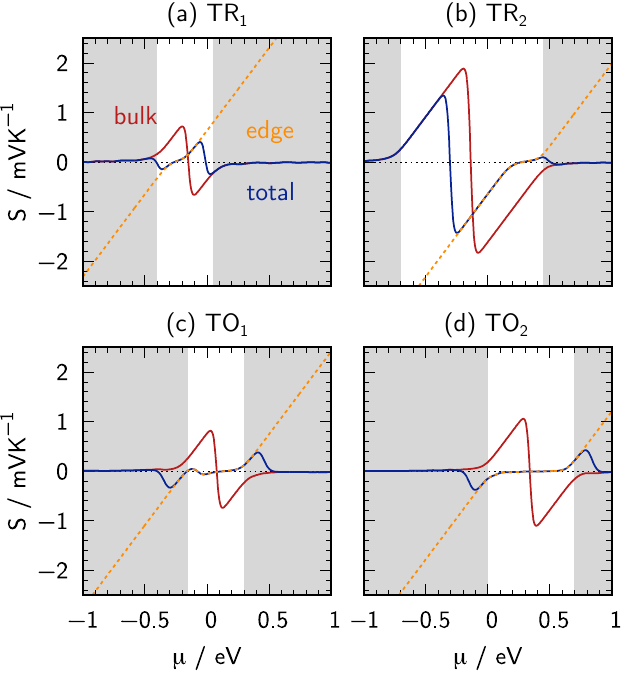}
\caption{The Seebeck coefficient $S$ of trivial (TR) phases (a) and (b) and topological (TO) phases (c) and (d) as a function of the Fermi energy $\mu$. The yellow, red and blue lines respectively represent edge, bulk, and total contributions. The corresponding $\lambda_\mathrm{v}$ and $\lambda_\mathrm{OS}$ parameters for all figures refer to Fig.~\ref{fig:bandstructure}(c). Gray (white) area indicates the continuum bulk (band gap) states. }
\label{fig:seebeck}
\end{figure}

We sample the thermoelectric coefficients for two trivial (TR$_1$ and TR$_2$) and two topological phases (TO$_1$ and TO$_2$). We provide both contributions from bulk states, edge states, and the total of them. It is shown that from experimental fitting, at room temperature, bulk monolayer bismuthene has relaxation time $\tau_{\rm{bulk}} = 15.24 \; \textrm{ps}$. In the edge states energy range, we have extremely long relaxation time, which is about two order of magnitude longer than bulk states~\cite{wang2018strongly}. In this analysis, we chose $\tau_{\rm{edge}} = 100\times \tau_{\rm{bulk}}$ for topological states and $\tau_{\rm{edge}} =\tau_{\rm{bulk}}$ for the trivial ones. Also, from molecular dynamics, in room temperature, monolayer buckled bismuthene has phonon thermal conductivity $\kappa_{ph} = 3.8\; \textrm{W/mK}$\cite{tauandkappa}.

In Fig. \ref{fig:sigma}, the bulk states only contribute to conductivity if the chemical potential lies outside the band gap (white area) for all phases. For edge contribution, it is also expected that in the absence of SOC effect, there is no edge state or carrier in the Fermi level, hence the edge contribution at $\mu=0$ is zero, as shown in TR$_1$ and TR$_2$. In trivial states, the edge states contribution is non-zero if the chemical potential lies below (TR$_1$) or above (TR$_2$) the Fermi level. For example, non-zero edge states below the Fermi level in TR$_1$ is caused by trivial edge state shown in fig. \ref{fig:edgebandstructure}(a). When the staggered potential is larger, the edge state below chemical potential will be suppressed, while the edge state above chemical potential becomes more relevant, as shown by the transition from TR$_1$ to TR$_2$. On the other hand, when the SOC effect is non-zero, the system undergoes a topological phase transition, the edge conductivity will contribute fully inside bulk band gap as shown in TO$_1$ phase. In TO$_2$ phase, the edge conductivity in the band gap becomes larger with the increases of $\lambda_{\textrm{os}}$ and $\lambda_{\textrm{v}}$. The band gap in that phase also increases due to larger $\lambda_\mathrm{v}$. The topological edge conductivity is larger by $100$ times than the bulk because we have set $\tau_\mathrm{edge}=100\times\tau_\mathrm{bulk}$.

\begin{figure}[tb]
\includegraphics[width=.95\columnwidth]{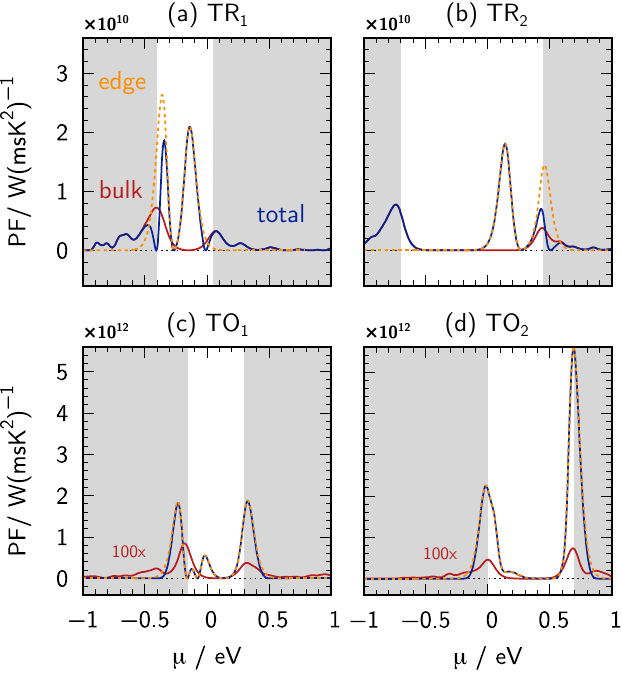}
\caption{The power factor (PF)  of trivial (TR) phases (a) and (b) and topological (TO) phases (c) and (d) as a function of the Fermi energy $\mu$. The yellow, red and blue lines respectively represent edge, bulk, and total contributions. The corresponding $\lambda_\mathrm{v}$ and $\lambda_\mathrm{OS}$ parameters for all figures refer to Fig.~\ref{fig:bandstructure}(c). Note that the multiplicators in (a) and (b) are $\times 10^{10}$ while in (c) and (d) are $\times 10^{12}$. In (c) and (d), the bulk contribution is deliberately multiplied by 100 times to show the $\mu$ dependence.   Gray (white) area indicates the continuum bulk (band gap) states.}
\label{fig:PF}
\end{figure}

\begin{figure}[tb]
\includegraphics[width=.95\columnwidth]{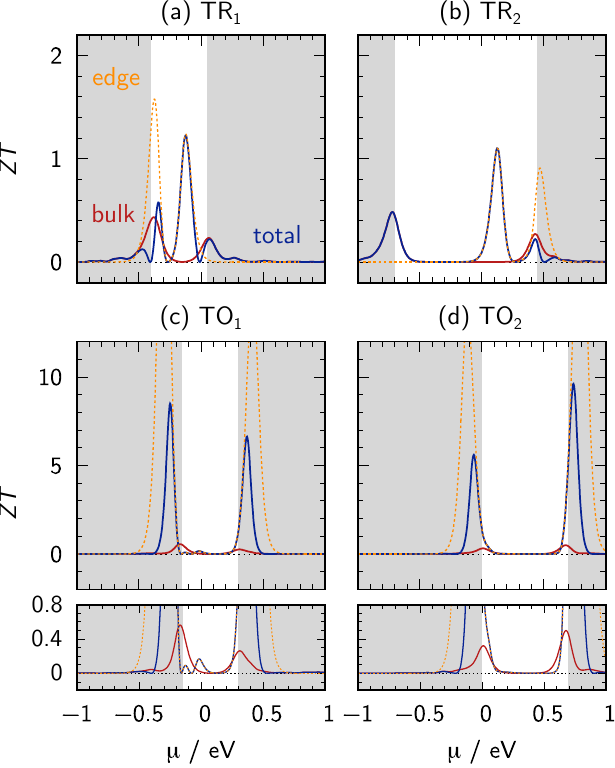}
\caption{Figure of merit \textit{ZT}  of trivial (TR) phases (a) and (b) and topological (TO) phases (c) and (d) as a function of the Fermi energy $\mu$. The yellow, red and blue lines respectively represent edge, bulk, and total contributions. The corresponding $\lambda_\mathrm{v}$ and $\lambda_\mathrm{OS}$ parameters for all figures refer to Fig.~\ref{fig:bandstructure}(c).  Gray (white) area indicates the continuum bulk (band gap) states.}
\label{fig:ZT}
\end{figure}

The electron thermal conductivity $\kappa_e$ follows the behavior of $\sigma$ indicating the good agreement with the Wiedemann-Franz law~\cite{ziman2001electrons}. Thus the same explanation in $\sigma$ behavior of TR and TO phases can also explain thermal conductivity of electron, as shown in Fig. \ref{fig:kappae}. 

The behavior of Seebeck coefficient $S$ of the edge tends to have the opposite sign from that of the bulk as shown in Fig. \ref{fig:seebeck}. These behaviors can be understood from the Mott formula, in which the Seebeck coefficient is proportional to the change of electrical conductivity, $S\propto -\frac{\partial_\mu\sigma(\mu)}{\sigma(\mu)}$. In the bulk states, $|S|$ is optimized near the charge neutrality $\mu=0$ because the conductivity changes the most rapidly at this point. The different sign of $S$ is related to the majority carrier of either electron or hole doping. 

Now we focus the analysis on the Seebeck coefficient of the trivial (TR) edge states. The trivial edge bands can be thought of as flat bands. The conductivity is localized only at the narrow region where the edge states are located [see Figs.~\ref{fig:sigma}(a) and (b)]. At large $\mu>0$ ($\mu<0$), $\partial_\mu\sigma$ picks a negative (positive) sign; as a result, $S$ becomes positive (negative). The same analysis can be applied to the TO states with the only difference being the bandwidth that extends over the whole band gap. In the band gap, $S$ of TO edge states become nearly zero.  As a result, the total Seebeck coefficient becomes smaller than that of the bulk.  

As a whole, the thermoelectric performance is represented by the power factor PF and the figure of merit \textit{ZT} in Figs. \ref{fig:PF} and \ref{fig:ZT}, respectively. The power factor in the TR phase indicates an enhancement due edge carrier by 2-3 times larger than its bulk states at hole doping for TR$_1$ and electron doping for TR$_2$, as shown in Figs.~\ref{fig:PF}(a) and \ref{fig:PF}(b), respectively. The power factor can be enhanced by 100 times (notice that multiplicator $\times 10^{12}$ instead of $10^{10}$ in Figs.~\ref{fig:PF}(c) and (d)) in the topological phase but with a caveat that the doping should be high enough to reach the degenerate states $|\mu|>\Delta/2$ as shown in Figs.~\ref{fig:PF}(c) and \ref{fig:PF}(d). This drastic difference comes from the contribution of electrical conductivity of edge states with magnitude 100 times larger than its bulk and also optimized $S$ from edge states at large $\mu$. More enhancement is possible for larger  SOC constant and staggered potential as shown in Fig.~\ref{fig:PF}(d).

A contrast phenomenon is shown in the figure of merit \textit{ZT} (Fig.~\ref{fig:ZT}). While we have the enhancement for TR phase [Fig.~\ref{fig:ZT}(a) and (b)], \textit{ZT} is not enhanced or even reduced in TO phase when the Fermi energy is inside the band gap. The reason for this is due to the tiny Seebeck coefficient in the band gap for TO phases while the enhancement from $\sigma$ is canceled by $\kappa_e$ thanks to the Wiedemann-Franz law. We note that for bismuthene $\kappa_\mathrm{ph}\ll \kappa_e$, thus our reasoning above holds. As the doping goes above the band gap, \textit{ZT} enhancement can be larger than $8$ and it is proportional with the strength of SOC [cf. Fig.~\ref{fig:ZT}(c) and (d)]. In short, topology does enhance the thermoelectric figure of merit, but not necessarily when the Fermi level is within bulk band gap~\cite{xu2017topological}. In 2D bismuthene with high on-site SOC, the contribution from topological edge states appears at large doping when the topological insulator becomes nearly metallic.

\section{Conclusion}
We have calculated of the electronic properties of bismuthene through the tight-binding method. The parameters of SOC and staggered potential can be tuned on-demand to show the topological phase transition and the resulting thermoelectric transport. Although topological edge states may enhance electrical conductivity, electron thermal conductivity and even the power factor, the Seebeck coefficient actually decreases in the band gap. This results in the decrease of \textit{ZT} in the low-doped topological insulators. At large doped TI near the degenerate states, \textit{ZT} of topological insulators can be larger than 8 thus useful for thermoelectrics. This doping is easily achievable via the gate voltages. The PF and \textit{ZT} increase when SOC is stronger. The \textit{ZT} enhancement can also be achieved even with trivial edge states. However, boundary scattering that is not considered in this work might hamper the transport properties. The analysis of this work can be applied to general 2D topological insulators. 

\begin{acknowledgments}
M.G. dan S.A.W. contributed equally to this work. We appreciate a fruitful discussion with Dr.~A.R.T. Nugraha. E.H.H. acknowledges financial support by the National Research Fund, Luxembourg under grants ATTRACT 7556175, CORE 13579612, and CORE 14764976. Numerical calculations have been done by using Mahameru HPC facilities provided by Indonesian Institute of Sciences (LIPI).
\end{acknowledgments}

\bibliographystyle{apsrev4-2}
\bibliography{ref}

\end{document}